\input phyzzx

\def\IR{{\hbox{{\rm I}\kern-.2em\hbox{\rm R}}}}
\def\IB{{\hbox{{\rm I}\kern-.2em\hbox{\rm B}}}}
\def\IN{{\hbox{{\rm I}\kern-.2em\hbox{\rm N}}}}
\def\IC{{\ \hbox{{\rm I}\kern-.6em\hbox{\bf C}}}}

\def\IZ{{\hbox{{\rm Z}\kern-.4em\hbox{\rm Z}}}}
\def\to{\rightarrow}

\def\underarrow#1{\vbox{\ialign{##\crcr$\hfil\displaystyle
{#1}\hfil$\crcr\noalign{\kern1pt
\nointerlineskip}$\longrightarrow$\crcr}}}
%

\def\ltorder{\mathrel{\raise.3ex\hbox{$<$}\mkern-14mu
             \lower0.6ex\hbox{$\sim$}}}
\def\lesssim{\mathrel{\raise.3ex\hbox{$<$}\mkern-14mu
             \lower0.6ex\hbox{$\sim$}}}


\font\ilis=cmbx10 scaled \magstep2
\font\Hugo=cmbx10 scaled \magstep3

\input phyzzx
\overfullrule=0pt
\tolerance=5000
\overfullrule=0pt
\twelvepoint

\twelvepoint

\titlepage
\vskip 2truecm

\centerline{\Hugo On the Reduced SU(N) Gauge Theory }
\centerline{\Hugo in the  Weyl-Wigner-Moyal Formalism\foot{In Memory of
Dr.
V\'{\i}ctor
Mart\'{\i}nez Oliv\'e.}}
\vglue-.5in
\author{ Hugo Garc\'{\i}a-Compe\'an\foot{Present address: {\it School of
Natural Sciences, Institute for Advanced Study, Olden Lane, Princeton NJ
08540, USA.} E-mail: compean@sns.ias.edu},
 Jerzy F. Pleba\'nski \foot{E-mail: pleban@fis.cinvestav.mx} and
Norma Quiroz-P\'erez\foot{E-mail: nquiroz@fis.cinvestav.mx}}
\medskip
\address{Departamento de F\'{\i}sica
\break  Centro de Investigaci\'on y de
Estudios Avanzados del IPN.
\break Apdo. Postal 14-740, 07000, M\'exico D.F., M\'exico.}
\bigskip

\abstract{ Weyl-Wigner-Moyal formalism is used to describe the
large-$N$ limit of reduced SU$(N)$ quenching gauge theory. Moyal
deformation of Schild-Eguchi action is obtained.}

\vskip 1truecm
\noindent



\endpage

Many years ago G. `t Hooft has shown that drastic simplification occurs in
the gauge structure of SU$(N_C)$ Quantum Chromodynamics (QCD), if one
takes the number of colors $N_C$ to be infinite$^1$. Large-$N$ limit, for
instance, contains genuine non-perturbative information of field theory at
the classical and quantum levels. Chiral symmetry breaking, mass gap and
color confinement are important non-perturbative features which persist in
this limit. Large-$N$ techniques have been also applied to describe mesons
and baryons in a complete picture$^2$ and to matrix model approach to 2D
quantum gravity and 2D string theory\foot{ For a recent review see
reference 3 and references therein.} when one looks for their
non-perturbative formulation. More recently large-$N$ technique has been
used to formulate the M(atrix) Theory$^4$ of zero D branes at the infinite
momentum frame.  This theory is by now a strong candidate to realize the
called $M$ Theory$^5$.  The power of these results enable one to study
some large-$N$ limits of various physical systems. For instance, some
two-dimensional integrable systems seem to be greatly affected in this
limit, turning out `more integrable'. Example of this `induced
integrability' occurs in the large-$N$ limit of SU$(N)$ Nahm
equations$^6$. Actually the origin of this drastic simplification in the
field equations is not well understood$^7$. The results presented in this
paper might will sheed some light about this mysterious limit.

The usual transition from SU$(N)$ gauge theory to the SU$(\infty)$ ones
involves the change of the Lie algebra su$(N)$ by the area-preserving
diffeomorphisms algebra, sdiff$(\Sigma)$, on a two-dimensional manifold
$\Sigma.$ The last one is an infinite dimensional Lie algebra.  As we make
only local considerations we assume the space $\Sigma$ to be a
two-dimensional simply connected and {\it compact} symplectic manifold
with local real coordinates $\{\tau, \sigma \}$. This space has a natural
local symplectic structure given by the local area form $\omega = d\sigma
\wedge d \tau$.  sdiff$(\Sigma)$ is precisely the Lie algebra associated
with the infinite dimensional Lie group, SDiff$(\Sigma)$, which is the
group of diffeomorphisms on $\Sigma$ preserving the symplectic structure
$\omega$, {\it i.e.} for all $g\in {\rm SDiff}(\Sigma)$, $g^*(\omega) =
\omega$.

Globally the symplectic form is defined by $\omega : T\Sigma \to T^*
\Sigma$ and inverse $\omega^{-1}:T^* \Sigma \to T \Sigma$.  Here $T\Sigma$
and $T^* \Sigma$ are the respective tangent and cotangent bundles to
$\Sigma.$ While the hamiltonian (or area preserving) vector fields are
${\cal U}_{H_a} = \omega^{-1}(dH_a)$ satisfying the sdiff$(\Sigma)$
algebra

$$[{\cal U}_{H_a},{\cal U}_{H_b}] = {\cal U}_{\{H_a,H_b\}_P}, \ \ \ {\rm
for \ all } \ (a\not = b), \eqno(1)$$ where $\{\cdot,\cdot \}_P$ stands
for the Poisson bracket with respect to $\omega$. Locally it can be
written as

$$\{H_a,H_b\}_P = \omega^{-1}(dH_a,dH_b) = \omega^{ij} \partial_i H_a
\partial_j H_b, \eqno(2)$$
where $\partial_i \equiv {\partial \over \partial \sigma^i}$, $(i=0,1),$
$\sigma^0 = \tau,$ $\sigma^1 = \sigma$ and $H_i= H_i(\vec{\sigma})=
H_i(\sigma^0,\sigma^1).$

The generators of sdiff$(\Sigma)$ are the hamiltonian vector fields ${\cal
U}_{H_a}$ associated to the hamiltonian functions $H_a$ given by

$$ {\cal U}_{H_a} = {\partial H_a \over \partial \sigma^0} {\partial \over
\partial \sigma^1} - {\partial H_a \over \partial \sigma^1} {\partial
\over \partial \sigma^0}.  \eqno(3)$$

On the other hand, in the Ref. 8, the Lie algebra su$(N)$ is defined in a
basis which appears to be very useful for our further considerations and
we going briefly review.

The elements of this basis are denoted by $L_{\vec m}$, $L_{\vec n}$,...,
etc., ${\vec m} = (m_1,m_2)$, ${\vec n} = (n_1,n_2)$,..., etc., and ${\vec
m}, \ {\vec n},...  \in I_N \subset {\bf Z} \times {\bf Z} - \{(0,0) \
{\rm mod} \ N{\vec q} \}$ where ${\vec q}$ is any element of ${\bf Z}
\times {\bf Z}$. The basic vectors $L_{\vec m}, \ {\vec m}\in I_N,$ are
the $N \times N$ matrices satisfying the following commutation relations

$$ [L_{\vec m}, L_{\vec n}] = {N \over \pi} Sin \big( {\pi \over N} {\vec
m} \times {\vec n} \big) L_{{\vec m} + {\vec n}}, \ \ {\rm mod} \ N{\vec
q},\eqno(4)$$
where ${\vec m} \times {\vec n} := m_1n_2 - m_2 n_1$.

Now we let $N$ tend to infinity. In this case $I_{\infty} \equiv I = {\bf
Z} \times {\bf Z} - \{(0,0)\}$ and the commutation relations (4) read

$$[L_{\vec m}, L_{\vec n}] = ({\vec m} \times {\vec n}) \ L_{{\vec m} +
{\vec n}}. \eqno(5)$$

Consider the complete set of periodic hamiltonian functions $ \{e_{\vec
m}\}_{{\vec m} \in I}$, $e_{\vec m} = e_{\vec m}(\vec{\sigma}) := {\rm
exp} \big[ i(m_1 \sigma^0 + m_2 \sigma^1) \big]$. One quickly finds that

$$ \{e_{\vec m}, e_{\vec n} \}_P = ({\vec m} \times {\vec n}) \ e_{{\vec
m} + {\vec n}}. \eqno(6)$$

Thus the mapping $F: L_{\vec m} \mapsto e_{\vec m}, \ \ {\vec m} \in I$,
defines the isomorphism

$$ {\rm su}(\infty) \cong {\rm the \ Poisson \ algebra \ on} \ \Sigma (=
T^2) \cong{\rm sdiff}(T^2), \eqno(7)$$
where $T^2$ is the 2-torus.

The algebra (4) is defined for the 2-torus $T^2$ but it can be extended to
a Riemann surface of genus $g \geq 1$ $\Sigma_g$ as has been shown by I.
Bars$^9$.  His argument is as follows: Consider the group SU$(N)$ with
$N=N_1 + \dots + N_g$ and define the set of $N\times N$ matrices,
$L_{\vec{m}_1}^{(N_1)} \oplus {\bf 1}_{N_2}\oplus \dots \oplus {\bf
1}_{N_g}$, ${\bf 1}_{N_1} \oplus L_{\vec{m}_2}^{(N_2)}\oplus \dots \oplus
{\bf 1}_{N_g}$, ${\bf 1}_{N_1} \oplus {\bf 1}_{N_2}\oplus \dots \oplus
L_{\vec{m}_g}^{(N_g)}$ where $L_{\vec{m}_k}^{(N_k)}$ and ${\bf 1}_{N_k}$
are $N_k \times N_k$ matrices (for $k=1, \dots, g),$ being the later the
unit matrix.

The generalization of (4) to a Riemann surface of genus $g$ is$^9$

$$ [L_{\vec{m}_1 \dots \vec{m}_g}, L_{\vec{n}_1 \dots \vec{n}_g}] = C_NSin
\bigg( \pi \sum_{i=1}^g {\vec{m}_i  \times \vec{n}_i\over
N_i} \bigg) L_{\vec{m}_1 + \vec{n}_1 \dots \vec{m}_g + \vec{n}_g}, \ \
{\rm mod} \ \big(N_1 \vec{q}_1, \dots N_g \vec{q}_g\big), \eqno(8)$$
where $(\vec{q}_1, \dots , \vec{q}_g) \in {\bf Z}^g
\times {\bf Z}^g.$ Here the generators $L_{\vec{m}_1 \dots \vec{m}_g}$ are
defined by

$$L_{\vec{m}_1 \dots \vec{m}_g}= \big(L_{\vec{m}_1}^{(N_1)}
\oplus {\bf 1}_{N_2}\oplus \dots \oplus {\bf 1}_{N_g}\big) \oplus
\big({\bf 1}_{N_1} \oplus L_{\vec{m}_2}^{(N_2)}\oplus \dots \oplus {\bf
1}_{N_g}\big) \oplus \dots \oplus \big({\bf 1}_{N_1} \oplus {\bf
1}_{N_2}\oplus \dots \oplus L_{\vec{m}_g}^{(N_g)}\big). \eqno(9)$$

It is shown in Ref. 9 that as one take the large-$N$ limit of (8) it
yields

$$ [L_{\vec{m}_1 \dots \vec{m}_g}, L_{\vec{n}_1 \dots \vec{n}_g}] =
\sum_{i=1}^g (\vec{m}_i \times \vec{n}_i) L_{\vec{m}_1 + \vec{n}_1 \dots
\vec{m}_g + \vec{n}_g}, \eqno(10)$$
which generalizes (5). The set of hamiltonian functions $e_{\vec{m}_1
\dots \vec{m}_g}$ associated with $L_{\vec{m}_1 \dots \vec{m}_g}$ are
defined by

$$ e_{\vec{m}_1 \dots \vec{m}_g} = {\rm exp}\bigg( i \sum _{i=1}^g
\vec{m}_i \cdot \vec{\sigma}_i \bigg) \eqno(11) $$
and satisfy the Poisson algebra$^9$

$$ \{e_{\vec{m}_1 \dots \vec{m}_g}, e_{\vec{n}_1 \dots \vec{n}_g}\} =
\sum_{i=1}^g (\vec{m}_i \times \vec{n}_i) e_{\vec{m}_1 + \vec{n}_1 \dots
\vec{m}_g + \vec{n}_g}. \eqno(12)$$

On the other hand, Bars using a series of basic results in reduced and
quenched large-$N$ gauge theories was able to derive the string theory
action in a particular gauge$^{10}$. In this derivation he used the above
mentioned area-preserving diffeomorphisms formalism on $T^2$\foot{
Large-$N$-limit was first studied by Hoppe in the context of membrane
physics$^{11}$. In that case $\Sigma = S^2$, the two-sphere.}.

A very similar (but different) approach was previously considered by
Fairlie, Fletcher and Zachos in the context of large-$N$ limit of
Yang-Mills theory in Ref. 8, and reviewed by Zachos in Ref. 12.  There it
was derived Nambu's action from the large-$N$ approach to Yang-Mills gauge
theory$^{13}$. In this derivation the quadratic Schild-Eguchi action for
strings$^{14}$ arose by the first time from a gauge theory. In what
follows we will take Bars approach$^{9,10}$ with quenched prescriptions by
Gross-Kitazawa$^{15}$.

The reduced SU$(N)$ gauge theory action is$^{9,10}$

$$ S_{red} = - {1 \over 4} \big({2 \pi \over \Lambda} \big)^d {N
\over
g^2_d(\Lambda)} Tr \big( {\cal F}_{\mu \nu}{\cal F}^{\mu \nu} \big),
\eqno(13)$$
where $d$ is the dimension of space-time space $M_d$, $g_d(\Lambda)$ is
the Yang-Mills
coupling constant in $d$ dimensions evaluated at certain cut-off $\Lambda$
and $Tr$ is an invariant bilinear form on the Lie algebra su$(N),$
${\cal F}_{\mu \nu}(x) = \partial_{\mu} A_{\nu}(x) -
\partial_{\nu}A_{\mu}(x) + i g_d[A_{\mu}(x),A_{\nu}(x)]$ with $x\in
M_d$, $\mu, \nu =
0,1,\dots , d-1$ and $A_{\mu}(x)$ is the usual Yang-Mills potential on
$M_d$.

>From now on we going follow Gross-Kitazawa paper$^{15}$. Thus the general
prescription to obtain the reduced scheme from Yang-Mills gauge theory is

$$ F_{\mu \nu} = [i{\cal D}_{\mu}, i{\cal D}_{\nu}] \Rightarrow ({\cal
F}_{\mu \nu})^i_j \equiv [a_{\mu}, a_{\nu}]^i_j, \eqno(14)$$
where $(\cdot)^i_j$ denotes an $N\times N$ matrix, ${\cal D}_{\mu}$ is the
covariant derivative with respect to the Yang-Mills potential $A_{\mu}$, $
i {\cal D}_{\mu}$ must be replaced by

$$ a_{\mu} = P_{\mu} + A_{\mu}, \eqno(15)$$
where $P_{\mu}$ is the quenched momentum which is a diagonal
matrix and $A_{\mu}$ is an $N\times N$ matrix gauge field at $x^{\mu} =
0$.

As quenched theory is a SU$(N)$ gauge theory, matrices $A_{\mu}$ must
satisfy a quenched gauge transformation

$$ A_{\mu} \to SA_{\mu} S^{\dag} + S [P_{\mu}, S^{\dag}], \eqno(16a)$$
$$ a_{\mu} \to S [a_{\mu}] S^{\dag}, \eqno(16b)$$
where $S$ is an unitary matrix. These transformations and the usual
quenched relations

$$ A_{\mu}(x) \equiv {\rm exp}\big( i P \cdot x \big) A_{\mu} {\rm exp}
\big(-i P\cdot x \big), \eqno(17a)$$

$$ S(x) \equiv {\rm exp}\big( i P \cdot x \big) S {\rm exp}
\big(-i P\cdot x \big), \eqno(17b)$$
where $P\cdot x \equiv P_{\mu} x^{\mu}$, lead to the usual gauge
transformation

$$A_{\mu}(x) \to S(x) A_{\mu}(x) S^{\dag}(x) + i S(x)
\partial_{\mu}S^{\dag}(x) \eqno(17c)$$
where $S: M_d \to {\rm SU}(N).$

The quanched euclidean Feynman integral is$^{15}$

$$ Z_N = \int \prod_{\mu} {\cal D} a_{\mu} f(a_{\mu}) exp\bigg( -
S_{red}(a)
\bigg). \eqno(18)$$
Condition (15) implies a further gauge invariant constraint on function
$f(a_{\mu})$ to be the quenched constraint between the eigenvalues of
$a_{\mu}$ and those of $P_{\mu}$ given by

$$ a_{\mu} = V_{\mu} P_{\mu} V^{\dag}_{\mu}, \eqno(19) $$
where $V_{\mu}$ diagonalizes $a_{\mu}$ for each ${\mu}$. At the quantum
level this quenched prescription besides the usual gauge fixing
terms involves an extra factor

$$ f(a) = \int \prod_{\mu} {\cal D}V_{\mu} \delta \big(a_{\mu} - V_{\mu}
P_{\mu} V^{\dag}_{\mu} \big) \eqno(20)$$
in the measure of the Feynman integral (18).


Now we will use the previous discussion on area preserving diffeomorphisms
to apply it to reduced large-$N$ gauge theory (for details see Refs.
8-13). Any solution of the SU$(N)$ reduced gauge theory equations coming
from the reduced action (13) can be written in the form

$$a_{\mu} = \sum_{{\vec m} \in I_N} a_{\mu}^{\vec m}(N) L_{\vec m}
\eqno(21)$$
where the set $\{L_{\vec m}\}$ satisfy algebra (4) for $T^2$ or (8) for
$\Sigma_g$ in its corresponding basis.

The object ${\cal F}_{\mu \nu} = [a_{\mu},a_{\nu}]$ can be written from
(4) and (21) as

$${\cal F}_{\mu \nu} = \sum_{{\vec m} \in I_N}\sum_{{\vec n} \in I_N}
a_{\mu}^{\vec m}(N) a_{\nu}^{\vec n}(N) {N \over \pi} Sin \bigg({\pi \over
N} {\vec m} \times {\vec n}\bigg) L_{\vec{m} + \vec{n}}, \ \ {\rm mod}\
N{\vec q}.
\eqno(22)$$
In the large-$N$ limit the above relation can be written equivalently in
terms of the hamiltonian function ${\cal F}^{\infty}_{\mu
\nu}(\vec{\sigma})$ in the basis $\{e_{\vec m}(\vec{\sigma})\}$

$$ {\cal F}^{\infty}_{\mu
\nu}(\vec{\sigma})  = \{ {\cal A}_{\mu}(\vec{\sigma)}, {\cal
A}_{\nu}(\vec{\sigma}) \}_P = \sum_{\vec{m},\vec{n}} a_{\mu}^{\vec m}
a_{\nu}^{\vec n}( {\vec m} \times {\vec n})e_{\vec{m} +
\vec{n}}(\vec{\sigma}), \eqno(23)$$
where ${\cal A}_{\mu}(\vec{\sigma}) = {\cal A}_{\mu}(\sigma^0,\sigma^1),$
$e_{\bf m}(\vec{\sigma})$ are the generator functions of sdiff$(T^2)$,
while $\{\sigma^0, \sigma^1 \}$ are the coordinates on the 2-torus $T^2$.
It is clear that the limit

$$ a_{\mu}^{\vec m} : = \lim_{N\to {\infty}} a_{\mu}^{\bf m} (N)
\eqno(24)$$
exists for every ${\vec m} \in I.$

Let $S^{\infty}_{red}$ be the action which is an $N \to \infty$ limit of
$S_{red}$. This limit can be obtained formally by the
substitutions$^{8-13}$

$$ {(2\pi)^4 \over N^3} Tr (\cdot \cdot \cdot ) \to - \int_{\Sigma} (\cdot
\cdot \cdot) d^2\sigma, \eqno(25a)$$

$$ a_{\mu} \to {\cal A}_{\mu}, \eqno(25b)$$

 $$[a_{\mu}, a_{\nu}] \to \{ {\cal A}_{\mu},{\cal A}_{\nu} \}_P.
\eqno(25c)$$
where $d^2 \sigma \equiv d\sigma^0 d \sigma^1.$

The large-$N$ limit of the reduced action $S_{red}$ is given by$^{9,10}$

$$ S^{\infty}_{red} = 4{(2\pi/\Lambda)^{d-4} \over g_d^2(\Lambda)}
\bigg({N \over 2 \Lambda} \bigg)^4 \int_{\Sigma} d^2 \sigma \ {\cal
F}^{\infty}_{\mu \nu}(\vec{\sigma}) {\cal F}^{\infty
\mu\nu}(\vec{\sigma}).  \eqno(26) $$

Using (2), Eq. (23) reads

$$ {\cal F}^{\infty}_{\mu \nu}(\vec{\sigma}) = \{ {\cal A}_{\mu}
(\vec{\sigma}), {\cal A}_{\nu}(\vec{\sigma}) \}_P = \omega^{ij} \partial_i
{\cal A}_{\mu}(\vec{\sigma}) \partial_j {\cal A}_{\nu} (\vec{\sigma}).
\eqno(27)$$

Bars has shown that action (26) turns out to be

$$ S^{\infty}_{red} \sim \int_{\Sigma} d^2 \sigma \ det \big(\partial_i
{\cal A} \cdot \partial_j {\cal A} \big) \eqno(28)$$
where ``$ \cdot $'' stands for the inner product $\partial_i {\cal A}
\cdot \partial_j {\cal A} \equiv \partial_i {\cal A}^{\mu} \partial_j
{\cal A}_{\mu}.$

The above action is a particular case of Polyakov's action with a flat
$d$-dimensional target space $M_d$

$$ S_{Pol}= \int_{\Sigma} d^2 \sigma \sqrt{-h} h^{ij} \partial_i {\cal
A}_{\mu}(\vec{\sigma}) \partial_j {\cal A}^{\mu}(\vec{\sigma}), \eqno(29)
$$
where the world-sheet metric $ h_{ij} = \partial_i {\cal A}_{\mu}
\partial_j {\cal A}^{\mu}$ is subject to the gauge condition $det(h) =
-1$.

Action (28) is also known as the Schild-Eguchi action$^{8,9,10,12,14}$ and
it is shown to be equivalent to Nambu's action.

Just as Bars shown, the corresponding path integral is

$$ Z_{\infty} = \sum_g \int {\cal D}m {\cal D} {\cal A}_{\mu} f({\cal A})
{\rm exp} \big( S_{red}^g({\cal A})\big) \eqno(30)  $$
which is the bosonic string amplitude  in the particular gauge
$det(h) = -1$, $m$ is the moduli which must be compatible with the
large-$N$ limit of SU$(N)$ and $f({\cal A})$ is the large-$N$ limit of
(20) in the basis $\{e_{\vec m}(\vec{\sigma}) \}.$

Now we attempt to derive Bars results using Weyl-Wigner-Moyal formalism in
quantum mechanics (see e.g. Ref. 16 and references therein). This
technique has been used to obtain some new solutions of Park-Husain
heavenly equation in self-dual gravity as well as to address some
relations between self-dual gravity and two-dimensional field
theories$^{17,18}$. Here we will apply this technique to rederive the
string
action.

{\it Weyl correspondence} ${\cal W}$ establishes a one to one
correspondence between some class of linear operators ${\cal B}$ acting on
Hilbert space ${\cal H} = L^2({\bf R})$ and the space of real smooth
functions $C^{\infty}(\Sigma, {\bf R})$ on the phase space manifold
$\Sigma$.
This correspondence is given by

$$ {\cal W}^{-1}: {\cal B} \to C^{\infty}(\Sigma, {\bf R}), \eqno(31)$$

$$ {\cal O}(\vec{\sigma};\hbar) \equiv {\cal W}^{-1}(\hat{\cal O}) :=
\int_{\infty}^{\infty} <\sigma - {\xi \over 2}|\hat{\cal O}|\sigma + {\xi
\over 2}> {\rm exp}\big( {i \over \hbar} \xi \tau\big) d\xi, \eqno(32) $$
for all $\hat{\cal O} \in {\cal B}.$ Of course ${\cal
O}(\vec{\sigma};\hbar) \in C^{\infty}(\Sigma, {\bf R}).$ The `inverse'
Weyl correspondence

$${\cal W}: C^{\infty}(\Sigma, {\bf R}) \to {\cal B}, \eqno(33)$$
is given by

$$\hat{\cal O} = {\cal W}({\cal O}(\vec{\sigma};\hbar)) := {1 \over (2
\pi)^2} \int_{\Omega \subset {\bf R}^2} \tilde{\cal O}(p,q) {\rm exp} [i(p
\hat{\sigma} + q
\hat{\tau})] dp dq \eqno(34) $$
and

$$ \tilde{\cal O} = \tilde{\cal O}(p,q) = \int_{\Sigma} {\cal
O}(\vec{\sigma};\hbar) {\rm exp}[-i(p \sigma^0 + q \sigma^1)] d^2 \sigma,
\eqno(35)$$
where $\tilde{\cal O}(p,q)$ is the Fourier transform of ${\cal
O}(\vec{\sigma};\hbar).$ Here the operators $\hat{\sigma}, \hat{\tau}$
satisfy the Heisenberg algebra and $\{p,q\}$ are the coordinates of the
Fourier dual space.

The Moyal-$\star$-product on $C^{\infty}(\Sigma; {\bf R})$ is defined by

$$ {\cal O}_i \star {\cal O}_j := {\cal O}_i {\rm exp} ({i {\hbar} \over
2} \buildrel{\leftrightarrow}\over{\cal P}){\cal O}_j, \ \ (i\not = j)
\eqno(36)$$
where $\buildrel {\leftrightarrow}\over {\cal P} := {\buildrel
{\leftarrow}\over{\partial} \over \partial \sigma^0} {\buildrel
{\rightarrow}\over{\partial} \over \partial \sigma^1} - {\buildrel
{\leftarrow}\over{\partial} \over \partial \sigma^1} {\buildrel
{\rightarrow}\over{\partial} \over \partial \sigma^0},$ ${\cal O}_i =
{\cal O}_i (\vec{\sigma};\hbar).$ The Moyal product is an associative
and non-commutative product.

With the above Moyal product definition and Eqs. (32), (34) and (35) it is
very easy to check that

$${\cal W}^{-1}(\hat{\cal O}_i \circ \hat {\cal O}_j ) = {\cal O}_i \star
{\cal O}_j \eqno(37) $$
where ``$\circ$'' stands for operator product in ${\cal B}.$ Using the
above results one can get the following relation

$${\cal W}^{-1} \big( {1 \over i \hbar} [\hat {\cal O}_i, \hat{\cal O}_j]
\big) = {1 \over i \hbar} ({\cal O}_i \star {\cal O}_j - {\cal O}_j \star
{\cal O}_i) \equiv \{{\cal O}_i,{\cal O}_j \}_M, \eqno(38)$$
where $[,]$ is the usual commutator and ${\cal W}^{-1}(\hat{\cal O}_i)
\equiv {\cal O}_i.$

Eqs. (36) and (38) lead to

$$\{{\cal O}_i,{\cal O}_j \}_M = {2 \over \hbar} {\cal O}_i Sin ({\hbar
\over 2} \buildrel {\leftrightarrow}\over {\cal P}) {\cal O}_j, $$

$$ = \sum_{k=0}^{\infty} { (-1)^k \over (2k+1)!} ({{\hbar}\over 2})^{2k}
\big( {\cal O}_i \buildrel {\leftrightarrow}\over {\cal P} {\cal O}_j
\big). \eqno(39)$$

Then one obtain that ${\cal W}^{-1}$ is a Lie algebra isomorphism

$${\cal W}^{-1} : \big({\cal B}, [,]\big) \to \big({\cal M}, \{\cdot,
\cdot \}_M \big), \eqno(40)$$
being ${\cal W}$ its genuine inverse map. $\big({\cal M},\{\cdot, \cdot
\}_M \big)$ is called {\it the Moyal algebra} which we abbreviate as
${\cal M}$.

Moyal algebra ${\cal M}$ is the {\it unique} ``quantum deformation'' of
Poisson algebras sdiff$(\Sigma)$$^{19}$. We represent the Moyal algebra
${\cal M}$ by sdiff$_{\hbar}(\Sigma),$ where $\hbar$ is the deformation
parameter.  In the limit $\hbar \to 0$ one recovers Poisson algebra,
$\lim_{\hbar \to 0}$ sdiff$_{\hbar}(\Sigma)=$ sdiff$(\Sigma).$ That is,
Weyl-Wigner-Moyal formalism also provide the corespondence with the
area-preserving formalism

$$\lim_{{\hbar}\to 0} {\cal O}_i \star {\cal O}_j = {\cal O}_i {\cal O}_j
\ \ {\rm and} \
\
\lim_{{\hbar} \to 0} \{{\cal O}_i, {\cal O}_j\}_M = \{{\cal
O}_i,{\cal O}_j\}_P. \eqno(41)$$

In the case when $\tau$ and $\sigma$ be the local coordinates of a
2-torus, {\it i.e.} $\Sigma = T^2.$ A basis of sdiff$_{\hbar}(\Sigma)$ is
given by

$$e_{\vec m}= e_{\vec m}(\vec{\sigma}) \equiv {\rm exp}[i(m_1 \sigma^0 +
m_2
\sigma^1)]. \eqno(42)$$
Introducing the last equation into (39) we get

$$\{e_{\vec m}(\vec{\sigma}),e_{\vec n}(\vec{\sigma}) \}_M = {2 \over
\hbar}
Sin \big[ {\hbar \over 2}({\vec{m} \times \vec{n}}) \big]
e_{\vec{m}+ \vec{n}}(\vec{\sigma}). \eqno(43)$$

Now we have a Fourier series instead of (34)

$${\cal W}\big({\cal O}(\vec{\sigma};\hbar)\big) = \hat{\cal O} = {1 \over
(2 \pi)^2} \sum_{\vec m} \tilde{\cal O}^{\vec m} e_{\vec m}(\vec{\sigma})
\eqno(44)$$
where $\tilde{\cal O}^{\vec m}$ is given by

$$\tilde{\cal O}^{\vec m}= \int_{T^2} {\cal O}(\vec{\sigma};\hbar) {\rm
exp} \big[ - i (m_1 \sigma^0 + m_2 \sigma^1) \big] d^2 \sigma. \eqno(45)$$

>From Eqs. (43) and (44) it is immediate to get

$$ \{ {\cal O}_i(\vec{\sigma};\hbar),{\cal O}_j(\vec{\sigma};\hbar) \}_M =
{1 \over (2 \pi)^4} \sum_{\vec{m}, \vec{n}} {2 \over \hbar} Sin \big[
{\hbar \over 2} ({\vec{m} \times \vec{n}}) \big] \tilde{\cal O}^{\vec m}
\tilde{\cal O}^{\vec n} e_{\vec{m} + \vec{n}}(\vec{\sigma}). \eqno(46)$$

Using the results of Ref. 9 one can immediate generalize the algebra (43)
defined on $T^2$ to a Riemann surface of genus $g$, $\Sigma_g$, as
follows:

First of all consider a set of hamiltonian functions (11) ${\cal O}_{i_1
\dots i_g}$ with the generalized Moyal product

$$ {\cal O}_{i_1 \dots i_g} \star {\cal O}_{j_1 \dots j_g} := {\cal
O}_{i_1 \dots i_g} {\rm exp} ({i {\hbar} \over 2}
\sum_{l=1}^g\buildrel{\leftrightarrow}\over{\cal P}_l){\cal O}_{j_1 \dots
j_g}, \ \ (i_k\not = j_k)  \eqno(47)$$
where $\buildrel {\leftrightarrow}\over
{\cal P}_l := {\buildrel {\leftarrow}\over{\partial} \over \partial
\sigma^0_l} {\buildrel {\rightarrow}\over{\partial} \over \partial
\sigma^1_l} - {\buildrel {\leftarrow}\over{\partial} \over \partial
\sigma^1_l} {\buildrel {\rightarrow}\over{\partial} \over \partial
\sigma^0_l}$ and where $\{\sigma^0_l, \sigma^1_l\}$ ($l=1, \dots ,g$) are
the coordinates of the $l-th$ 2-torus. These coordinates satisfy
$\{\sigma^0_l, \sigma^1_{l'}\}_P = \delta_{ll'}$. With the above
definitions we can find that the Moyal product for the Riemann surface is

$$e_{\vec{m}_1 \dots \vec{m}_g}\star e_{\vec{n}_1 \dots \vec{n}_g} =
{\rm exp}\Bigg( i {\hbar \over 2} \sum_{i=1}^g \vec{m}_i \times \vec{n}_i
\Bigg)
e_{\vec{m}_1 + \vec{n}_1 \dots \vec{m}_g + \vec{n}_g}. \eqno(48) $$
Thus the generalization of (43) is

$$\{e_{\vec{m}_1 \dots \vec{m}_g}(\vec{\sigma}),e_{\vec{n}_1 \dots
\vec{n}_g}(\vec{\sigma}) \}_M = {2 \over \hbar} Sin \bigg( {\hbar \over
2} \sum_{i=1}^g\vec{m}_i \times \vec{n}_i \bigg)
e_{\vec{m}_1 + \vec{n}_1 \dots \vec{m}_g + \vec{n}_g}(\vec{\sigma}).
\eqno(49)$$

Take the deformation parameter $\hbar$ to be

$$ \hbar = {2 \pi \over \sum_{i=1}^g N_i}= { 2 \pi\over N} \eqno(50)$$
and comparing algebras (8) and (49) one can establish an isomorphism
between both algebras. Thus the formalism used in this paper can be easily
extended from a 2-torus to a general Riemann surface. From now on we
will restrict  ourselves to work on $T^2$ with the immediate
generalization to $\Sigma_g$ when it be required.

Now we begin from the operator $\hat{\rm SU}(N)$-vauled reduced gauge
theory. Here $\hat{\rm SU}(N)$ is the Lie group of linear unitary
operators acting on the Hilbert space ${\cal H}=L^2({\bf R})$. Let ${\cal
B} = \hat{\rm su}(N)$ be the corresponding Lie algebra of the
anti-self-dual operators in $L^2({\bf R})$.  The reduced action now takes
the operator form

$$S^{(q)}_{red}:= - {1 \over 4} ({2 \pi \over \Lambda})^d { N \over
g_d^2(\Lambda)} {\rm Tr} \big [\hat{\cal F}_{\mu \nu} \hat{\cal F}^{\mu
\nu} \big], \eqno(51)$$
where $\hat{\cal F}_{\mu \nu} \in \hat{\rm su}(N),$ `Tr' is the sum over
diagonal elements with respect to an orthonormal basis $\{ |\psi_j>\}_{j
\in {\bf {\bf N}}}$ in $L^2({\bf R})$

$$ <\psi_j| \psi_k> = \delta_{jk}, \ \ \ \ \ \sum_j |\psi_j><\psi_j| =
\hat I. \eqno(52)$$

The general prescription $(14,15)$ can be written now as

$$ \hat {F}_{\mu \nu} = \bigg([i\hat {\cal D}_{\mu}, i\hat{\cal
D}_{\nu}]\bigg) \Rightarrow (\hat{\cal F}_{\mu \nu})^i_j \equiv [\hat{\cal
A}_{\mu}, \hat{\cal A}_{\nu}]^i_j, \eqno(53)$$
where $ \hat{\cal A}_{\mu} = \hat{P}_{\mu} + {\hat A}_{\mu},$
$\hat{P}_{\mu}, \hat{A}_{\mu} \in \hat{\rm su}(N).$

Action (51) can be written in the basis $\{ | \psi>_i \}_{i\in {\bf N}}$
as

$$ S^{(q)}_{red}= -{1 \over 4} ({2 \pi \over \Lambda})^d { N \over
g_d^2(\Lambda)}\sum_j <\psi_j| \bigg \{ [\hat{\cal A}_{\mu}, \hat{\cal
A}_{\nu}][\hat{\cal A}^{\mu}, \hat{\cal A}^{\nu}] \bigg \} |\psi_j>.
\eqno(54)$$

We can rewrite Eq. (54) in the following form

$$ S^{(q)}_{red} = {1 \over 4} ({2 \pi \over \Lambda})^d { N \over
g_d^2(\Lambda)} \hbar^2\sum_j <\psi_j| \bigg \{ ({1 \over i
\hbar})[\hat{\cal A}_{\mu}, \hat{\cal A}_{\nu}]({1 \over i
\hbar})[\hat{\cal A}^{\mu}, \hat{\cal A}^{\nu}] \bigg \} |\psi_j>.
\eqno(55)$$

Now we arrive at the point where the Weyl-Wigner-Moyal formalism$^{16-18}$
can be applied.  By the {\it Weyl correspondence} ${\cal W}^{-1}$ one gets
the real function on the flat surface $\Sigma$ defined as follows (see Eq.
(32))

$$ {\cal A}_{\mu}(\vec{\sigma};\hbar) := {\cal W}^{-1}({\cal A}) = \int_{-
\infty}^{+ \infty} <\sigma - {\xi\over 2}| \hat{\cal A}_{\mu} | \sigma +
{\xi \over 2}> {\rm exp} \big( {i \tau \xi \over{\hbar}} \big) d\xi.
\eqno(56)$$

The action (55) transforms after a computation

$$S^{(q)}_{red} = {1 \over 4}\big({2\pi \over \Lambda}\big)^{d-1}{N \over
\Lambda} {\hbar \over g_d^2(\Lambda)} \int_{\Sigma} d^2 \sigma \ {\cal
F}^{(M)}_{\mu \nu}(\vec{\sigma};\hbar) \star {\cal F}^{(M) \mu
\nu}(\vec{\sigma};\hbar), \eqno(57)$$
where

$${\cal F}^{(M)}_{\mu \nu}(\vec{\sigma};\hbar) = \{ {\cal
A}_{\mu}(\vec{\sigma};\hbar),{\cal A}_{\nu}(\vec{\sigma};\hbar)  \}_M.
\eqno(58)$$


Now we focus in the quanching prescription within the WWM-formalism. First
we observe from the operational relation $\hat{\cal A}_{\mu} =
\hat{P}_{\mu} + \hat{A}_{\mu}$ and the definition (56) that

$$ {\cal A}_{\mu}(\vec{\sigma};\hbar) = P_{\mu}(\vec{\sigma};\hbar) +
A_{\mu}(\vec{\sigma};\hbar). \eqno(59)$$

Operator valued quenched theory is still a $\hat{\rm SU}(N)$ gauge theory
and so operator-valued quenched gauge transformations ($16a,b$) and
($17a,b$) still hold. Using WWM-formalism we found that  $(16a)$ and
$(16b)$ are now

$$A_{\mu}(\vec{\sigma};\hbar) \to S(\vec{\sigma};\hbar) \star
A_{\mu}(\vec{\sigma};\hbar) \star S^{- \buildrel{\star}\over{1}}
(\vec{\sigma};\hbar) + S(\vec{\sigma};\hbar)\star \{
P_{\mu}(\vec{\sigma};\hbar), S^{- \buildrel{\star}\over{1}}
(\vec{\sigma};\hbar)\}_M, \eqno(60a) $$

$$a_{\mu}(\vec{\sigma};\hbar) \to S(\vec{\sigma};\hbar)\star
a_{\mu}(\vec{\sigma};\hbar) \star S^{- \buildrel{\star}\over{1}}
(\vec{\sigma};\hbar), \eqno(60b)$$
where now $S(\vec{\sigma};\hbar) \in {\cal W}^{-1} \big(\hat{\rm
SU}(N)\big) \equiv {\rm SU}(N)_{\star}$. The later is an infinite
dimensional
Lie Group which is defined as$^{18}$

$$ {\rm SU}(N)_{\star} := \{ S= S(\vec{\sigma};\hbar) \in
C^{\infty}(\Sigma)
/ S^{- \buildrel{\star}\over{1}} (\vec{\sigma};\hbar)\star
S(\vec{\sigma};\hbar) = 1, $$

$$ S(\vec{\sigma};\hbar) \star S^{-
\buildrel{\star}\over{1}}(\vec{\sigma};\hbar)= 1; \ \ \
\bar{S}(\vec{\sigma};\hbar)
= S^{- \buildrel{\star}\over{1}} (\vec{\sigma};\hbar) \}. \eqno(61)$$
where `bar' stands for complex conjugation.
Relations ($17a,b$) are written in WWM-formalism as

$$ A_{\mu} (x,\vec{\sigma};\hbar) \equiv {\rm exp}_{\star}\bigg({i\over
\hbar}P_{\mu}(\vec{\sigma};\hbar)x^{\mu} \bigg) \star
A_{\mu}(\vec{\sigma};\hbar)\star {\rm exp}_{\star}\bigg(-{i\over
\hbar}P_{\mu}(\vec{\sigma};\hbar)x^{\mu} \bigg), \eqno(62a)$$

$$ S(x,\vec{\sigma};\hbar) \equiv {\rm exp}_{\star}\bigg({i\over
\hbar}P_{\mu}(\vec{\sigma};\hbar)x^{\mu} \bigg) \star
S(\vec{\sigma};\hbar)\star {\rm exp}_{\star}\bigg(-{i\over
\hbar}P_{\mu}(\vec{\sigma};\hbar)x^{\mu} \bigg), \eqno(62b)$$
where ${\rm exp}_{\star}\bigg({i \over \hbar} P_{\mu}(\vec(\sigma;\hbar)
x^{\mu}\bigg)$ is defined as$^{20,18}$

$$ {\rm exp}_{\star}\bigg({i\over \hbar} P_{\mu}x^{\mu}\bigg):=
\sum_{n=0}^{\infty} {1 \over n!} ({ix
\over
\hbar})^n P \star \dots \star P,  \ \ (n-{\rm times}), \eqno(63)$$
where $ P(\vec{\sigma};\hbar)= \sum_{n=0}^{\infty} \hbar^n
P_n(\vec{\sigma})$$^{21}$.

Thus substituting Eqs. ($62a,b$) into ($60a,b$) and using (38) one can
prove that the gauge transformation still holds

$$A_{\mu}(x, \vec{\sigma};\hbar) \to S(x,\vec{\sigma};\hbar)\star
A_{\mu}(x, \vec{\sigma};\hbar)\star S^{-
\buildrel{\star}\over{1}}(x,\vec{\sigma};\hbar) +
S(x,\vec{\sigma};\hbar)\star \partial_{\mu} S^{-
\buildrel{\star}\over{1}}(x,\vec{\sigma};\hbar), \eqno(64)$$
where $S(x,\vec{\sigma};\hbar)$ can be seen also as a smooth real function
$S:M_d\times \Sigma \to {\bf R}.$

At the quantum level the quenched constraint over the eigenvalues of
$a_{\mu}$ and $P_{\mu}$ (19) translates to

$$a_{\mu}(\vec{\sigma};\hbar) = V_{\mu}(\vec{\sigma};\hbar) \star
P_{\mu}(\vec{\sigma};\hbar) \star V_{\mu}^{- \buildrel{\star} \over
{1}}(\vec{\sigma};\hbar). \eqno(65)$$

Thus the quenched prescription on the eigenvalues is translated to the
functional relation (65). This functional prescription can be implemented
in the Feynman integral as

$$ f\big(a_{\mu}(\vec{\sigma};\hbar)\big) = \int \prod_{\mu} {\cal
D}V_{\mu}(\vec{\sigma};\hbar) \delta \bigg(a_{\mu}(\vec{\sigma};\hbar) -
V_{\mu}(\vec{\sigma};\hbar)\star
P_{\mu}(\vec{\sigma};\hbar) \star
V^{-\buildrel{\star}\over{1}}_{\mu}(\vec{\sigma};\hbar) \bigg),
\eqno(66)$$
where one have to integrate over the infinite dimensional subspace ${\cal
L} \subset C^{\infty}(\Sigma)$ such the the $V$'s satisfy conditions (61).
It is an easy matter to see, with the help of Eqs. (41), that one takes
the limit $\hbar \to 0$ our quenched functional prescription corresponds
exactly with large-$N$ quenched prescription of Bars$^{9,10}$. For
instance  Eq. (66) corresponds with $f({\cal A})$ function which appears
in Eq. (30).

Now we going to study the reduced quenched action (57). Expressing the
Moyal bracket $\{ {\cal A}_{\mu},{\cal A}_{\nu} \}_M$ as a deformed
Poisson bracket in the spirit of Strachan$^{22}$

$$ \{ {\cal A}_{\mu},{\cal A}_{\nu} \}_M = \omega^{ij} \partial_i {\cal
A}_{\mu}(\vec{\sigma};\hbar) \star \partial_j {\cal A}_{\nu}
(\vec{\sigma};\hbar), \eqno(67) $$
we define a $\star$-deformed ``world-sheet metric''

$$ \buildrel{\star}\over{h}_{ij} \equiv \partial_i {\cal
A}(\vec{\sigma};\hbar) \star \partial_j {\cal A} (\vec{\sigma};\hbar).
\eqno(68) $$
Now assume that this metric will transform as

$$ \buildrel{\star}\over{h}^{ij} = \omega^{lm}
\buildrel{\star}\over{h}_{lm} \omega^{ij}. \eqno(69) $$

The quenched quantum action (57) now reads proportional

$$S^{(q)}_{red} \sim \int_{\Sigma} d^2 \sigma \buildrel{\star} \over
{h}^{ij} \partial_i {\cal A}^{\mu}(\vec{\sigma};\hbar) \star \partial_j
{\cal A}_{\mu}(\vec{\sigma};\hbar). \eqno(70)$$
This is the Moyal deformation of Schild-Eguchi action.

It can be shown that the above action can be derived from the ``Moyal
deformation of Polyakov's action'' (or quantum Polyakov)

$$S^{(q)}_{Pol} \sim \int_{\Sigma} d^2 \sigma \sqrt{- \buildrel{\star}
\over {h}} \buildrel{\star} \over {{h}^{ij}} \partial_i {\cal
A}^{\mu}(\vec{\sigma};\hbar) \star \partial_j {\cal
A}_{\mu}(\vec{\sigma};\hbar), \eqno(71)$$
where $\buildrel{\star}\over{h} \equiv {\rm det}(\buildrel{\star} \over
{{h}^{ij}} )$ and det$: GL(2,{\bf R}) \to C^{\infty}(\Sigma,{\bf R}).$

Following Strachan$^{21}$ we assume that
${\cal A}_{\mu}(\vec{\sigma};\hbar)$ is an analytic function in $\hbar$
{\it i.e.}

$${\cal A}_{\mu}(\vec{\sigma};\hbar) = \sum_{k =0}^{\infty} \hbar^k {\cal
A}^{(k)}_{\mu}(\vec{\sigma}). \eqno(72)$$

Taking the $\hbar \to 0$ limit one can see that our quenched quantum
action (70)
can be reduced to the action (29) for the zero component ${\cal
A}^{(0)}_{\mu}(\vec{\sigma})$ of (72), {\it i.e.}

$$ \lim_{{\hbar} \to 0} S^{(q)}_{red} = \int_{\Sigma} h^{ij}
\partial_i{\cal A}^{(0) \mu} (\vec{\sigma}) \partial_j {\cal
A}^{(0)}_{\mu}(\vec{\sigma}) = S^{\infty}_{red}. \eqno(73)$$

Finally we find that Feynman integral (30) is still valid in the context
of WWM-formalism but besides to the usual moduli and gauge constraints
there is an additional constraint on the space of ${\cal A}$'s.  It is
that one must perform the functional integration on those ${\cal A}$'s
which admit Moyal deformation. Of course the limit $\hbar \to 0$
reproduces exactly Bars result (30).

At the present paper we have constructed another way to obtain the
large-$N$ limit of reduced gauge action. We begin from the $\hat{\rm
su}(N)$ reduced gauge theory and apply Weyl-Wigner-Moyal formalism and
then the $\hbar \to 0$ limit. We find the Moyal deformation of the
quenched gauge theory and in particular the Moyal deformation of
Schild-Eguchi action, which in the $\hbar \to 0$ gives the Schild-Eguchi
action. This resul is valid for any underlying Riemann surface $\Sigma_g$
with ($g\geq1$) for the phase space. Based in the recent result by
Bars$^9$ we generalize the Moyal algebra for its appropriate use. Also we
shown that the usual quenched prescription on the momenta$^{15}$ is
translate to a functional ones (65,66). On the other hand we think that
our formulation is more appropriate for the discussion of the large-$N$
limit and quantum mechanics analogies of the quenched gauge theory. This
is because in Moyal algebra the limiting process ($\hbar \to 0$)is well
defined unlike the limit $N\to \infty$ in matrix models. Moreover from a
theoretical point of view it is more elegant to deal with a deformation of
Poisson-Lie algebra than with the matrix algebras which depend on discrete
parameter $N$. Following the arguments by Fairlie$^{23}$ concerning the
the suitability of application of Moyal brackets to $M$ Theory, we believe
that our approach can be applied$^{24}$ straighforward to the reduced
matrix model$^{25}$ formulated recently in the context of the
M(atrix)-Theory. It would be interesting to find the relation of our
results and those found by Bars in Ref. (26) concerning SU$(N)$ gauge
theory on discrete Riemann surfaces.  Although our method
 needs an {\it explicit} Lie algebra homomorphism $\Psi: {\rm su}(N) \to
\hat{\rm su}(N),$ which is not an easy matter, however explicit examples
are given in the search of some solutions of self-dual Einstein equations
for the Lie algebras su$(2)$ and sl$(2)$$^{6,17,18}$. This means that our
method at least works for the case $N=2$. The consideration for higher
values of $N$ will be consider in a forthcoming paper.

Our results can be summarized in the next commuting diagram:

\bigskip

$$
\def\mapright#1{\smash{
\mathop{\longrightarrow}\limits^{#1}}}

\matrix{
&&\hidewidth{\Psi}\hidewidth& & \cr
&\pmatrix{{\rm su}(N)~{\rm Reduced}\cr{\rm Gauge}\cr
{\rm Action}\cr}&\mapright{}& \pmatrix{\hat{\rm su}(N)~{\rm Reduced}\cr
{\rm Gauge} \cr{\rm Action}\cr}\cr
&&&\cr
\hidewidth{ N \to \infty}\hidewidth&\downarrow{}&&\downarrow{}&
\hidewidth{\cal W}^{-1} \hidewidth\cr
&&&\cr
&\pmatrix{{\rm sdiff}(\Sigma) \cong {\rm su}(\infty)\cr {\rm Reduced~Gauge}\cr
{\rm Action}\equiv {\rm string~action}\cr} & \leftarrow{} &
\pmatrix{{\rm Moyal~Deformation}\cr {\rm of~Reduced}\cr
{\rm Gauge~Action}\cr} & \cr
&&\hidewidth{\hbar \to 0}\hidewidth& & \cr}
$$

\bigskip

Finally the problem if the above results can be reproduced and generalized
by using the geometrical framework of deformation quantization geometry
used in Refs. 22 and 27 is still open.

\vskip 1truecm

\leftline{\ilis Acknowledgements}

We are greatly indebted to Profs. C.K. Zachos and M. Przanowski for useful
observations and comments, to Professor I. Bars por pointing out Ref. 26
and to Iliana Carrillo for mathematical discussions.  In addition, thanks
are due to CONACyT under the program {\it Programa de Posdoctorantes:
Estancias Posdoctorales en el Extranjero para Graduados en Instituciones
Nacionales (1996-1997)} and the Academia Mexicana de Ciencias under the
program: {\it Estancias de Verano para Investigadores J\'ovenes.}

\vskip 2truecm
\centerline{\Hugo References}

\item{1.} G. `t Hooft, Nucl. Phys. B {\bf 72} (1974) 461.

\item{2.} E. Witten, Nucl. Phys. B {\bf 160} (1979) 57.

\item{3.} P. Ginsparg and G. Moore, ``Lectures on 2D gravity and 2D
string Theory, TASI lectures 1992, Eds. J. Harvey and J. Polchinski (World
Scientific 1993).

\item{4.} T. Banks, W. Fischler, S.H. Shenker and L. Susskind,
``$M$-Theory as a Matrix Model: A Conjecture'', Phys. Rev. D {\bf 55}
(1997) 5112, hep-th/9610043; T. Banks, ``The State of Matrix Theory'',
hep-th/9706168.

\item{5.} E. Witten, ``String Theory Dynamics in Various Dimensions'',
Nucl. Phys. B {\bf 443} (1995) 85, hep-th/9603124; J.H. Schwarz, ``The
Power of $M$-Theory'', Phys. Lett. B {\bf 367} (1996) 97, hep-th/9510086.

\item{6.} R.S. Ward, Phys. Lett. {\bf B 234} (1990) 81; J. Geom. Phys.
{\bf 8} (1992) 317; H. Garc\'{\i}a-Compe\'an and J.F. Pleba\'nski, ``On
the Weyl-Wigner-Moyal Description of SU$(\infty)$ Nahm Equations'', to
appear in Phys. Lett. A, hep-th/9612221.

\item{7.} E. Witten, J. Geom. Phys. {\bf 8} (1992) 327.

\item{8.} D.B. Fairlie, P. Fletcher and C.K. Zachos, J. Math. Phys. 31
(1990) 1088.

\item{9.} I. Bars, ``Strings and Matrix Models On Genus $g$ Riemann
Surfaces'', hep-th/9706177.

\item{10.} I. Bars, ``Strings From Reduced Large-$N$ Gauge Theory Via
Area-Preserving Diffeomorphisms'', Preprint USC-90/HEP-12,
IASSNS-HEP-90/21, (1990); ``Area Preserving Diffeomorphisms as a Bridge
Between Strings and Large-$N$ QCD'', Proceedings, Strings 1990 pp.
202-209; ``Strings and Large-$N$ QCD'', Proceedings {\it Beyond the
Standard Model} Ed. K. Milton, (1991) pp. 67-72.

\item{11.} J. Hoppe, M.I.T. Ph. D. Thesis (1982); Constraints Theory and
Relativistic Dynamics- Florence 1986, G. Longhi and L. Lusanna eds. p.
267, World Scientific, Singapore (1987); Phys. Lett. B {\bf 215} (1988)
706.

\item{12.} C. Zachos, ``Hamiltonian Flows, SU$(\infty)$, SO$(\infty)$,
Usp$(\infty)$ and Strings, in {\it Differential Geometric Methods in
Theoretical Physics}, Eds. L.L. Chau and W. Nahm, Plenum Press, New York, 1990.

\item{13.} E.G. Floratos, J. Iliopoulos and G. Tiktopoulos, Phys. Lett.
{\bf B217} (1989) 285.

\item{14.} A. Schild, Phys. Rev. D {\bf 16} (1977) 1722; T. Eguchi, Phys.
Rev. Lett. {\bf 44} (1980) 126.

\item{15.} D.J. Gross and Y. Kitazawa, Nucl. Phys. B {\bf 206} (1982)
440.

\item{16.} J.F. Pleba\'nski, M. Przanowski, B. Rajca and J. Tosiek, Acta
Phys. Pol. B {\bf 26} (1995) 889; J.F. Pleba\'nski, M. Przanowski and J.
Tosiek, ``Weyl-Wigner-Moyal-Formalism II. The Moyal Bracket, Acta Phys.
Pol. B {\bf 27} (1996) 1961.

\item{17.} J.F. Pleba\'nski and M. Przanowski, Phys. Lett. A {\bf 212}
(1996) 22; J.F. Pleba\'nski M. Przanowski and H. Garc\'{\i}a-Compe\'an,
``From Principal Chiral Model to Self-dual Gravity'', Mod. Phys. Lett. {\bf
A11} (1996) 663.

\item{18.} H. Garc\'{\i}a-Compe\'an, J. F. Pleba\'nski and M.  Przanowski,
``Further remarks on the chiral model approach to self-dual gravity'',
Phys. Lett. A {\bf 219} (1996) 249.

\item{19.} P. Fletcher, ``The Uniqueness of the Moyal Algebra'', Phys.
Lett. B {\bf 248} (1990) 323.

\item{20.} K. Takasaki, J. Geom. Phys. {\bf 14} (1994) 332;
hep-th/9305169.

\item{21.} I.A.B. Strachan, Phys. Lett. B {\bf 283} (1992) 63.

\item{22.} I.A.B. Strachan, ``The Geometry for Multidimensional
Integrable Systems'', J. Geom. Phys. (1996), to appear.

\item{23.} D.B. Fairlie, ``Moyal Brackets in $M$-Theory'', hep-th/9707190.

\item{24.} H. Garc\'{\i}a-Compe\'an and L. Palacios, to appear.

\item{25.} N. Ishibashi, H. Kawai, Y. Kitazawa and A. Tsuchiya, ``A Large
$N$ Reduced Model as Superstring'', Nucl. Phys. B {\bf 498} (1997) 467,
hep-th/9612115.

\item{26.} I. Bars, ``SU$(N)$ Gauge Theory and Strings on Discrete Riemann
Surfaces'', USC-90-HEP-20 (1990).

\item{27.} H. Garc\'{\i}a-Compe\'an, J.F. Pleba\'nski and M. Przanowski,
``Geometry Associated with the Self-dual Yang-Mills and the Chiral Model
Approaches to Self-dual Gravity'', hep-th/9702046, submitted to J. Phys. A.

\endpage

\end